\documentclass[12pt]{article}

\usepackage{geometry}
\usepackage[ruled,vlined]{algorithm2e}
\usepackage{clipboard}
\newclipboard{quotes}
\usepackage{amsfonts}
\usepackage{amsmath}
\usepackage{amssymb}
\usepackage{amsthm}
\usepackage{array}
\usepackage{bm}
\usepackage{booktabs}
\usepackage{caption}
\usepackage{color}
\usepackage{graphicx}
\usepackage{hyperref}
\usepackage{multirow}
\usepackage[authoryear,semicolon]{natbib}
\usepackage{physics}
\usepackage{siunitx}
\usepackage{standalone}
\usepackage{subfigure}
\usepackage{tikz}
\usepackage[normalem]{ulem}
\usepackage{url}

\newcolumntype{x}[1]{>{\centering\arraybackslash\hspace{0pt}}p{#1}}
\newcommand{\E}{\text{E}}

\newcommand{\Var}{\text{Var}}

\newcommand{\cov}{\text{cov}}
\newcommand{\diag}{\text{diag}}

\newcommand{\gammavec}{\boldsymbol{\gamma}}
\newcommand{\betavec}{\boldsymbol{\beta}}

\newcommand{\etavec}{\boldsymbol{\eta}}
\newcommand{\muvec}{\boldsymbol{\mu}}

\newcommand{\Phivec}{\boldsymbol{\Phi}}

\newcommand{\Sigmavec}{\boldsymbol{\Sigma}}

\newcommand{\fvec}{\mathbf{f}}

\newcommand{\hvec}{\mathbf{h}}

\newcommand{\svec}{\mathbf{s}}

\newcommand{\xvec}{\mathbf{x}}
\newcommand{\zvec}{\mathbf{z}}

\newcommand{\Gau}{\text{Gau}}

\newcommand{\Cvec}{\mathbf{C}}

\newcommand{\Mvec}{\mathbf{M}}

\newcommand{\new}{\textrm{new}}

\let\originalleft\left
\let\originalright\right
\renewcommand{\left}{\mathopen{}\mathclose\bgroup\originalleft}
\renewcommand{\right}{\aftergroup\egroup\originalright}
\renewcommand{\arraystretch}{1.2}

\long\def\symbolfootnote[#1]#2{\begingroup%
\def\thefootnote{\fnsymbol{footnote}}\footnote[#1]{#2}\endgroup}

\begin{document}

\def\spacingset#1{\renewcommand{\baselinestretch}%
{#1}\small\normalsize} \spacingset{1.5}


  \title{\textbf{\Large{Warped Gradient-Enhanced Gaussian Process Surrogate Models for Exponential Family Likelihoods with Intractable Normalizing Constants}}}
  \author{\normalsize{Quan Vu,
     Matthew T. Moores,
    and Andrew Zammit-Mangion} \\
    \normalsize{School of Mathematics and Applied Statistics, University of Wollongong}
    }
  \date{}
  \maketitle

\begin{abstract}

Markov chain Monte Carlo methods for exponential family models with intractable normalizing constant, such as the exchange algorithm, require simulations of the sufficient statistics at every iteration of the Markov chain, which often result in expensive computations. Surrogate models for the likelihood function have been developed to accelerate inference algorithms in this context. However, these surrogate models tend to be relatively inflexible, and often provide a poor approximation to the true likelihood function. In this article, we propose the use of a warped, gradient-enhanced, Gaussian process surrogate model for the likelihood function, which jointly models the sample means and variances of the sufficient statistics, and uses warping functions to capture covariance nonstationarity in the input parameter space. We show that both the consideration of nonstationarity and the inclusion of gradient information can be leveraged to obtain a surrogate model that outperforms the conventional stationary Gaussian process surrogate model when making inference, particularly in regions where the likelihood function exhibits a phase transition.  We also show that the proposed surrogate model can be used to improve the effective sample size per unit time when embedded in exact inferential algorithms. The utility of our approach in speeding up inferential algorithms is demonstrated on simulated and real-world data.

\end{abstract}

\noindent%

\spacingset{1.5} 

\section{Introduction}\label{sec:intro}

Methods for statistical inference usually require the likelihood function to be evaluated pointwise, up to an unknown normalizing constant. However, many important exponential family models have an intractable likelihood that cannot be evaluated, but that can be easily simulated from. 
In the case of the Potts model \citep{potts1952} used for image analysis, and the exponential random graph model \citep[ERGM;][]{frank1986ergm} used for social network analysis, the likelihood function features a phase transition where, on a small region of the parameter space, the model behavior changes rapidly from one phase (known as the ordered phase) to another phase (known as the disordered phase). This property makes inference with these models even more challenging. 

A growing body of literature is concerned with computational methods for inference with models that have intractable likelihoods. For example, pseudo-marginal methods \citep{beaumont2003estimation,andrieu2009pseudo} make use of an unbiased estimate of the likelihood function, while in approximate Bayesian computation \citep[ABC;][]{tavare1997inferring,pritchard1999population}, summary statistics are used to compare simulated pseudo-data at given parameter values to observed data. One of the most popular approaches involves the use of Markov chain Monte Carlo (MCMC) with auxiliary variables. 
Two algorithms in this class include that introduced by \cite{moller2006efficient}, and the exchange algorithm \citep{murray2006mcmc}.
These MCMC methods require simulation of pseudo-data from the likelihood at each iteration of the Markov chain. In practice, Gibbs or Swendsen--Wang (SW) algorithms \citep{swendsen1987nonuniversal} are used for simulating the sufficient statistics. Even these algorithms can be computationally expensive when the data dimension is large, rendering inference infeasible in many applications. 

To make inference for these models computationally tractable, surrogate likelihoods are often employed to approximate the true likelihood function. Such methods can speed up inference, as they do not require expensive simulations of the sufficient statistics at every iteration. For example, \cite{boland2018efficient} used a deterministic function to emulate ratios of normalizing constants, while \cite{moores2020scalable} used a deterministic function to emulate the sufficient statistics. 
An attractive way for constructing surrogate models for the likelihood function is through Gaussian processes, which are flexible, probabilistic models. Gaussian process emulators were proposed in the context of computer experiments for modeling computationally expensive functions \citep[e.g.,][]{sacks1989design, kennedy2000predicting}. Gaussian process emulators were subsequently used as surrogate models in approximate Bayesian computation by \cite{meeds2014gps, wilkinson2014accelerating, jarvenpaa2018gaussian} and \citet{jarvenpaa2020parallel}. \cite{drovandi2018accelerating} and \cite{park2020function} employed Gaussian process surrogate models for facilitating Bayesian computation in an MCMC context.

Typically, Gaussian process surrogate models are constrained to be stationary. However, if the likelihood function undergoes a phase transition, the sufficient statistics can abruptly change with small changes in the input parameters at the transition; this sudden change in behavior is synonymous with nonstationarity when modeling using stochastic processes. In Section \ref{sec:potts} we show that using a stationary Gaussian process surrogate model for the sufficient statistics may lead to large errors when emulating said sufficient statistics. The surrogate model is often used directly, instead of the true likelihood function, in an MCMC algorithm, resulting in an inexact-approximate algorithm. In such cases, inferential accuracy heavily depends on the accuracy of the surrogate model.

In this paper we build a surrogate model to emulate the sufficient statistics for making computationally-efficient inference with exponential family models that have an intractable normalizing constant.  However, to rectify the aforementioned problems, we propose using a nonstationary Gaussian process to emulate the sufficient statistics, specifically one based on the warped Gaussian process model introduced by \citet{azm2019deep} and \citet{vu2021modeling}. The main contribution here, over these and related works that also consider nonstationary Gaussian process models \citep[e.g.][]{jarvenpaa2018gaussian,aushev2020likelihoodfree} is the incorporation of gradient information in our model \cite[see][for a review]{laurent2019overview}. This modification leads to a multivariate Gaussian process that \emph{jointly} models the means and the variances of the sufficient statistics. We show that gradient-enhanced nonstationary Gaussian process surrogate models offer a large improvement over both univariate stationary, and nonstationary, Gaussian process models when emulating the sufficient statistics, particularly in the vicinity of phase transitions. We illustrate the use of our surrogate model in importance sampling \citep{everitt2017bayesian,vihola2020is} and delayed-acceptance MCMC \citep{christen2005markov,sherlock2017da}, which target the exact posterior distribution over the parameters. We show that our proposed methodology may be used to good effect in both the complete-data setting and the incomplete-data setting.

The remainder of the article is organized as follows. In Section 2, we present our general approach for modeling the sufficient statistics to approximate the likelihood of exponential family models with intractable normalizing constant. In Section 3, we introduce the gradient-enhanced nonstationary Gaussian process surrogate models for the sufficient statistics, and detail the algorithms that make use of the surrogate models for inference. In Section 4, we demonstrate the use of the surrogate models on three data sets. Section 5 concludes.

\section{Background}

In this section, we define the models our approach is suitable for, and detail the likelihood function that we approximate with surrogate models in Section 3.

\subsection{Intractable Likelihood}
In this paper we consider models for which the likelihood function can be written in the following, exponential family, form,
\begin{equation}\label{eq:expLike}
p(\zvec \mid \betavec) = \frac{\exp{\betavec^T \svec(\zvec)}}{\mathcal{C}(\betavec)},
\end{equation}
where the normalizing constant is given by
\begin{equation}\label{eq:normConst}
\mathcal{C}(\betavec) = \sum_{\zvec \in \mathcal{Z}} \exp\left\{\betavec^T \svec(\zvec)\right\},
\end{equation}
$\zvec = (z_1, \dots, z_N)^T$ are the observed data, $\svec(\zvec) = \left(s^{(1)}(\zvec), \dots, s^{(D)}(\zvec)\right)^T$ are the sufficient statistics, and $\betavec = (\beta^{(1)}, \dots, \beta^{(D)})^T$ are the natural parameters.
When the set of  all possible observed data, $\mathcal{Z}$, is large, the computational cost of evaluating the sum \eqref{eq:normConst} becomes infeasible. There are models of this form that will benefit from  the methodology developed in this work. These include the Potts model and the related autologistic model \citep{besag1974spatial} that we discuss in Sections \ref{sec:potts} and \ref{sec:autologistic}, respectively, and the exponential random graph model \citep[ERGM, e.g.,][]{Robins_2007}.

\subsection{Approximate Likelihood}
We follow the approach of \cite{price2018bayesian}, and approximate the computationally intractable likelihood in \eqref{eq:expLike} using a Bayesian synthetic likelihood. Specifically, we approximate the intractable likelihood as a multivariate normal distribution of the sufficient statistics, with mean $\muvec(\betavec)$ and covariance $\Sigmavec(\betavec)$. This yields the synthetic likelihood function $\tilde{p}(\zvec \mid \betavec) = \mathcal{N}\left(\svec(\zvec) ; \muvec(\betavec), \Sigmavec(\betavec) \right)$. Note that if the sufficient statistics are highly non-Gaussian, there are more robust synthetic likelihood approaches that one can use \citep[e.g.,][]{an2020robust, Frazier_2021}. We further assume that the sufficient statistics in $\svec(\zvec)$ are mutually independent, that is, we let $\Sigmavec(\betavec) = \diag(\{ \sigma^{2 (d)}, d = 1,\dots, D \})$, so that
\begin{equation}\label{eq:syn_like}
\tilde{p}(\zvec \mid \betavec) = \prod_{d=1}^{D} \mathcal{N}(s^{(d)}(\zvec); \mu^{(d)}(\betavec), \sigma^{2 (d)}(\betavec)).
\end{equation}

To evaluate our synthetic likelihood function $\tilde{p}(\zvec \mid \betavec^{*})$ for any $\betavec^{*}$,  we build surrogate models for the mean functions $ \{ \mu^{(d)}(\betavec^{*}) \}_d $ and variance functions $ \{ \sigma^{2 (d)}(\betavec^{*}) \}_d$. We first simulate pseudo-data (using the SW algorithm) at a fixed set of parameters for fitting the surrogate model. Specifically, consider a fixed set of $p$ parameter values $\{\betavec_1, \dots, \betavec_p\}$. For each $\betavec_j, j = 1,\dots,p$, we generate $q$ simulations of the sufficient statistics $\{ \svec_{j, 1}, \dots, \svec_{j, q} \}$, where $\svec_{j, k} = (s_{j,k}^{(1)}, \dots, s_{j,k}^{(D)})^T$. Then, we obtain the sample means and the sample variances of the simulations at $\betavec_j$, that is, we compute
\begin{align}\label{eq:sample}
\begin{split}
m^{(d)}(\betavec_j) &= \frac{1}{q} \sum_{k=1}^{q}  s^{(d)}_{j, k}, \\
v^{(d)}(\betavec_j) &= \frac{1}{q - 1} \sum_{k=1}^{q}  (s^{(d)}_{j, k} - m^{(d)}(\betavec_j) )^2,
\end{split}
\end{align}
for $j = 1, \dots, p,$ and $d = 1, \dots, D$. The sample means $\{m^{(d)}(\betavec_j)\}_{j,d}$ and the sample variances $\{v^{(d)}(\betavec_j)\}_{j,d}$ are treated as (noisy) observations of the true means $\{\mu^{(d)}(\betavec_j)\}_{j,d}$ and variances $\{\sigma^{2 (d)}(\betavec_j)\}_{j,d}$, respectively. 

We note that, from \eqref{eq:expLike}, for $d = 1,\dots,D$,
\begin{align}\label{eq:derivative_property}
\begin{split}
\mu^{(d)}(\betavec) &= \E_{\betavec}(s^{(d)}(\zvec)) = \frac{\partial}{\partial \beta^{(d)}} \log \mathcal{C}(\betavec), \\
\sigma^{2 (d)}(\betavec) &= \Var_{\betavec}(s^{(d)}(\zvec)) = \frac{\partial^2}{\partial \beta^{(d)^2}} \log \mathcal{C}(\betavec),
\end{split}
\end{align}
which implies $\sigma^{2 (d)}(\betavec) = \frac{\partial}{\partial \beta^{(d)}} \mu^{(d)}(\betavec)$. This property motivates us to model the means and variances jointly in our surrogate model. Such models are often referred to as gradient-enhanced models \citep{laurent2019overview}.


\subsection{Prior Distribution}\label{sec:priors}

In this paper we employ independent, bounded uniform priors for the model parameter(s) $\betavec$, largely for computational convenience and to facilitate comparison with other related techniques that also employ uniform priors  \citep[e.g.][]{moller2006efficient,Everitt2012networks,lyne2015russian,jarvenpaa2020parallel}. Other priors should be considered by practitioners that take into account the context of their application. For example, in the case of the (one-parameter) Potts model described in Section~\ref{sec:potts}, it is often reasonable to assume that neighboring variables (pixels) in a lattice are positively correlated. A prior distribution that excludes negative values of the inverse temperature parameter $\beta$ is therefore reasonable with this model. In practice, there will also be a value of $\beta$, $\beta_{crit}$ say, beyond which realizations of the $k$-state Potts model will have fewer than $k$ unique labels with high probability. For this reason, it can be beneficial to put an upper bound on $\beta$, or at least penalize large values by using an exponential prior for $\beta$; this penalized complexity prior would be similar to those discussed by \cite{simpson2017penalising}. 

In many applications, there is also ample prior information available. For example, in the case of Landsat data, such as those considered in the example of Section~\ref{sec:hidden_potts}, satellite imagery is available from 1972 to present \citep{wulder2022fifty}, and the historical data could be used to construct a prior distribution. In these cases, calibrated log-normal, gamma, truncated normal, or scaled beta distributions are all suitable candidates. Expert elicitation can also be used to construct informative priors \citep{french2022soft}. Prior information is particularly useful in our context as it identifies regions of the parameter space where one should put most effort in  constructing a representative Gaussian process surrogate model (discussed in the following sections). 

\section{The Surrogate Model}\label{sec:surrogate_model}

In Section \ref{sec:gp_surrogate_model}, we introduce the gradient-enhanced nonstationary Gaussian process surrogate model, which uses both deformation functions and gradient information to improve the fit to the surrogate means. In Section \ref{sec:evaluate_surrogate}, we describe approaches for evaluating the surrogate synthetic likelihood at arbitrary parameter values using the fitted surrogate model. Section \ref{sec:inference_surrogate} presents a few ways with which one could use the surrogate synthetic likelihood for both inexact-approximate and exact-approximate Bayesian inference.

\subsection{Gaussian Process Surrogate Models}\label{sec:gp_surrogate_model}

In this section, we introduce the Gaussian process surrogate models for the sufficient statistics. The Gaussian process surrogates use the observed sample means $\{m^{(d)}(\betavec_j)\}_{j,d}$ and sample variances $\{v^{(d)}(\betavec_j)\}_{j,d}$ from \eqref{eq:sample} to emulate the true means and variances at any $\betavec^{*}$. 

It is important that simulator ``noise'' is accounted for in surrogate models \citep[e.g.,][Chap.~5]{gramacy2020surrogates}. This is relevant for our surrogate models, since we build them using sample (and not exact) means and variances. We therefore model each $m^{(d)}(\betavec_j)$ and $v^{(d)}(\betavec_j)$ as a noisy observation of the (true) surrogate mean $\mu^{(d)}(\betavec)$ and surrogate variance $\sigma^{2 (d)}(\betavec)$, respectively. Since the sample mean and the sample variance are asymptotically normally distributed around the mean and variance of the sufficient statistics, respectively, we let
\begin{align}\label{eq:data_model}
\begin{split}
m^{(d)}(\betavec_j) &\equiv \mu^{(d)}(\betavec_j) + \epsilon^{(d)}_{\mu j}, \\
v^{(d)}(\betavec_j) &\equiv \sigma^{2 (d)}(\betavec_j) + \epsilon^{(d)}_{\sigma j},
\end{split}
\end{align}
for $j = 1, \dots, p,$ and $d = 1, \dots, D$, where $\epsilon^{(d)}_{\mu j} \sim \mathcal{N}(0, \tau^{2 (d)}_{ \mu j})$ and $\epsilon^{(d)}_{\sigma j} \sim \mathcal{N}(0, \tau^{2 (d)}_{\sigma j})$ are Gaussian noise terms with variances $\tau^{2 (d)}_{ \mu j}$ and $\tau^{2 (d)}_{ \sigma j}$, respectively. Here, we assume that the noise terms are independent, but these could also be modeled as more general Gaussian processes \citep[e.g., ][Chap.~10]{gramacy2020surrogates}. 
For simplicity, here we estimate the variances $\tau^{2 (d)}_{ \mu j}$ and $\tau^{2 (d)}_{ \sigma j}$ under the assumption that the underlying sufficient statistics are Gaussian. Hence, we set 
$$
\tau^{2(d)}_{\mu j} = \frac{v^{(d)}(\betavec_j)}{q} \qquad \textrm{and} \qquad \tau^{2(d)}_{\sigma j} = \frac{2 v^{2(d)}(\betavec_j)}{q-1}.
 $$
Equation \eqref{eq:data_model} does not respect nonnegativity of the variance parameter; this was not found to be a problem in practice, and more complicated models could be considered (at some computational cost) if needed.

Note that although one can obtain both the sample means and variances from the observations (see \eqref{eq:sample}), one may choose to build just one surrogate model: one for the surrogate means or one for the surrogate variances. For example, \cite{moores2020scalable} built a surrogate model for the variances and used the mean--variance relationship given in \eqref{eq:derivative_property} to find the surrogate means by integration. \cite{park2021bayesian} built a surrogate model for the means, and fixed the surrogate covariance matrices to be equal to the nearest empirical covariance matrices. One could also employ coupled mean-variance Gaussian process models in this context \citep[e.g., Chapter 10 of][]{gramacy2020surrogates}; however, these do not take advantage of the mean-variance relationship given in \eqref{eq:derivative_property}.

In what follows we consider different Gaussian process surrogate models: first, a stationary Gaussian process surrogate model for the means, and then a nonstationary Gaussian process surrogate model for the means. Finally, we  build a novel joint Gaussian process surrogate for the means and variances that takes advantage of the relationship in \eqref{eq:derivative_property}. 


\subsubsection{Stationary Gaussian Process Surrogate}

We first consider a stationary Gaussian process for modeling the means of the sufficient statistics. In this first case, independently, we let
\begin{equation}\label{eq:stat_gp}
\mu^{(d)}(\cdot) = \tilde g^{(d)}(\cdot) = b^{(d)}(\cdot) + g^{(d)}(\cdot), \quad d = 1, \dots, D,
\end{equation}
where $b^{(d)}(\cdot)$ is the trend, and $g^{(d)}(\cdot)$ is a zero-mean stationary Gaussian process. This model serves as a baseline, against which we will evaluate the more sophisticated models we discuss next.

\subsubsection{Nonstationary Gaussian Process Surrogate}\label{sec:nonstat}

\begin{figure}
	\centering
	
	\includegraphics[width=0.8\textwidth]{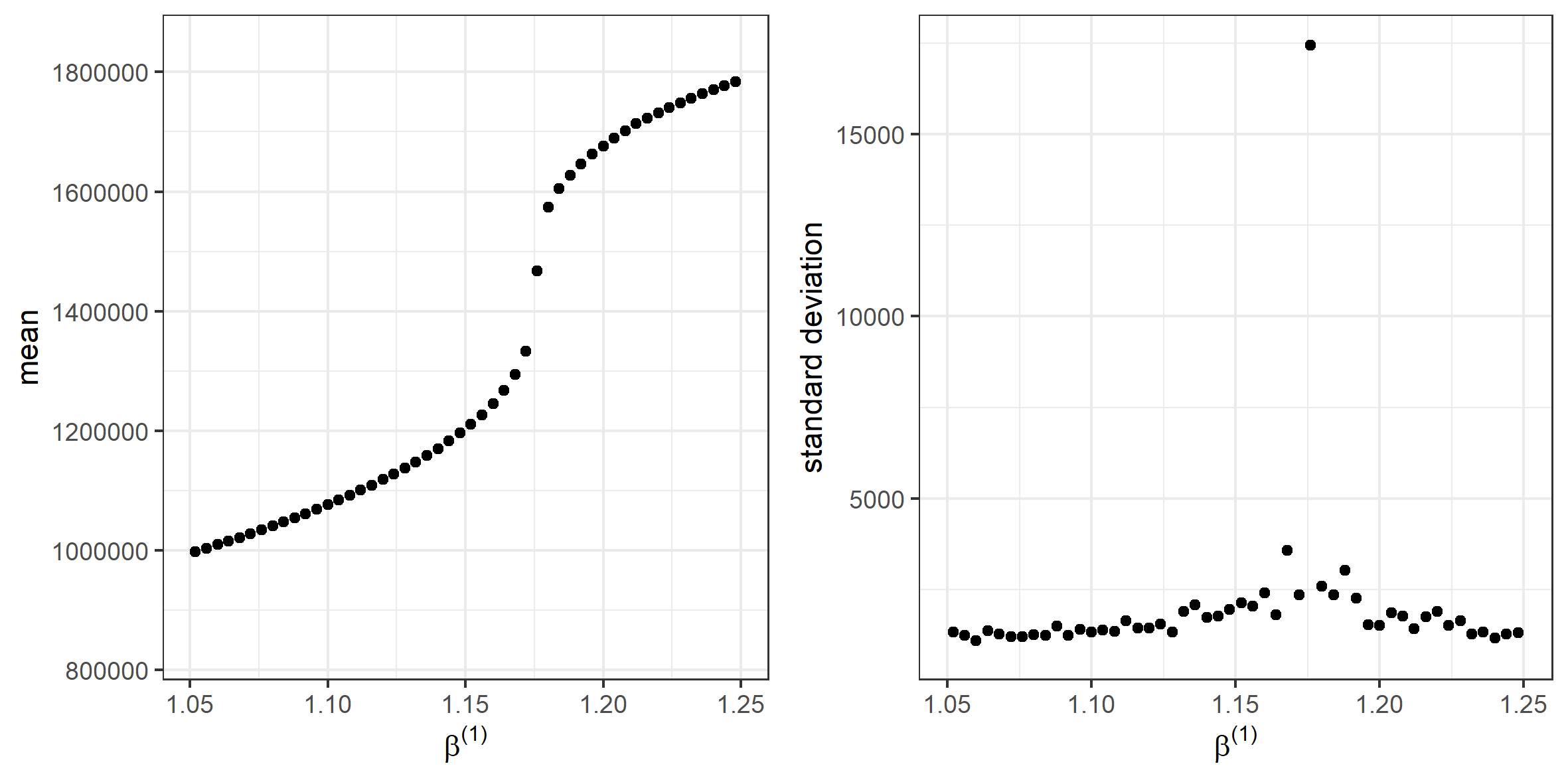}

	\caption{
	Empirical means (left panel) and standard deviations (right panel) of the simulated sufficient statistics for different values of $\beta^{(1)}$ for the Potts model on a $1000 \times 1000$ image with 5 labels.
	}
	\label{fig:potts_summary_stats}
\end{figure}

Phase transition is a property in many models with intractable likelihoods, such as the Potts model and the ERGM. Figure \ref{fig:potts_summary_stats} shows the sample means and sample standard deviations of the sufficient statistic of the Potts model (the sum of all neighbor pairs with the same label) for a $1000 \times 1000$ image with 5 labels (see Section \ref{sec:potts} for details). When the parameter $\beta^{(1)}$ is between 1.15 and 1.2, the mean of the sufficient statistic changes value rapidly with small changes in the parameters; this is in contrast to other regions in the parameter space where the rate of change is relatively slow. This region in the parameter space where the mean of the sufficient statistic changes rapidly is often called a phase transition.

As we show in Section \ref{sec:potts}, a stationary Gaussian process surrogate model cannot adequately capture the nonstationary behavior of the means of the sufficient statistics at the phase transition. We therefore also consider a nonstationary Gaussian process for the surrogate means. There are several approaches that could be adopted to model nonstationarity; here, we use the deformation (or warping) approach \citep{sampson1992nonparametric} where nonstationarity is obtained via a deformation (or warping) of the parameter space. The warping function $\fvec(\cdot)$ maps the parameter space onto a new domain on which a stationary Gaussian process is modeled, which induces a nonstationary Gaussian process on the original domain. We construct a flexible warping function $\fvec(\cdot)$ by linking several simple warping functions (such as basis functions) through composition, as detailed by \citet{azm2019deep}. Warping of the parameter space is a natural way forward when modeling the means of the sufficient statistics, as the warping function can stretch the parameter space when these undergo a phase transition and shrink the parameter space in other regions, in such a way that on the warped space it would be reasonable to model the process as stationary.

Our nonstationary model for the means of the sufficient statistics is given by
\begin{equation}\label{eq:nonstat_gp}
\mu^{(d)}(\cdot) = \tilde g^{(d)}(\cdot) = b^{(d)}(\cdot) + g^{(d)}(\fvec^{(d)}(\cdot)),
\end{equation}
where $b^{(d)}(\cdot)$ is the trend, $g^{(d)}(\cdot)$ is a zero-mean stationary Gaussian process, and $\fvec^{(d)}(\cdot)$ is a deformation/warping function given by $\fvec^{(d)}(\cdot) \equiv \fvec^{(d)}_n \circ~ \cdots~ \circ \fvec^{(d)}_1(\cdot)$ where each $\fvec^{(d)}_i(\cdot) = \Phivec^{(d)}(\cdot) \etavec_i^{(d)}$ is constructed using basis functions $\Phivec^{(d)}(\cdot)$ and weights $\etavec^{(d)}_i$, that need to be estimated. See \cite{azm2019deep} for more details on the basis functions and estimation.

\subsubsection{Gradient-Enhanced Nonstationary Gaussian Process Surrogate}

As shown in \eqref{eq:derivative_property}, the variance of the sufficient statistic, $\sigma^{2 (d)}(\betavec)$, is the derivative of the mean of the statistic, $\mu^{(d)}(\betavec)$, with respect to the $d^{\text{th}}$ dimension of $\betavec$. This motivates us to use a gradient-enhanced Gaussian process to jointly model the means and the variances of the sufficient statistics, to ultimately improve the fit to the surrogate means \citep{riihimaki2010gaussian, laurent2019overview}. This joint model is expected to improve the quality of the fit to the means of the sufficient statistics as, in this case, the sample variances are also informative on the means of the sufficient statistics.

We model the means of the sufficient statistics using nonstationary Gaussian processes as in \eqref{eq:nonstat_gp}. We then model the variances as the derivatives of the respective surrogate means. Therefore, our joint model of the mean $\mu^{(d)}(\cdot)$ and the variance $\sigma^{2 (d)}(\cdot)$ is a bivariate Gaussian process \citep{banerjee2003directional},
\begin{equation}\label{eq:ge_nonstat_gp}
\begin{pmatrix}
\mu^{(d)}(\cdot) \\ \sigma^{2 (d)}(\cdot)
\end{pmatrix} =
\begin{pmatrix}
\tilde g^{(d)}(\cdot) \\ \tilde h^{(d)}(\cdot)
\end{pmatrix} = 
\begin{pmatrix}
b^{(d)}(\cdot) \\ c^{(d)}(\cdot)
\end{pmatrix} + 
\begin{pmatrix}
g^{(d)}(\fvec^{(d)}(\cdot)) \\ h^{(d)}(\fvec^{(d)}(\cdot))
\end{pmatrix},
\end{equation}
where $b^{(d)}(\cdot)$ is the trend, $c^{(d)}(\cdot) = \frac{\partial}{\partial \beta^{(d)}} b^{(d)}(\cdot) $, $g^{(d)}(\cdot)$ is a zero-mean stationary Gaussian process, $\fvec^{(d)}(\cdot)$ is a deformation function, and $h^{(d)}(\cdot) = \frac{\partial}{\partial \beta^{(d)}} g^{(d)}(\cdot) $. This joint Gaussian process is a special case of the multivariate nonstationary Gaussian process with shared warping function proposed by \cite{vu2021modeling}.

The cross-covariance matrix function of the joint Gaussian process in \eqref{eq:ge_nonstat_gp} depends on the covariance function of $g^{(d)}(\cdot)$, which we denote by $K^{(d)}(\cdot, \cdot)$. The cross-covariance matrix function of the joint Gaussian process in \eqref{eq:ge_nonstat_gp}, which we denote by $\Cvec^{(d)}(\cdot,\cdot)$, is then given by
$$
\Cvec^{(d)}(\betavec_j, \betavec_l) = \begin{pmatrix}
  \cov(\tilde g^{(d)}(\betavec_j), \tilde g^{(d)}(\betavec_l)) & \cov(\tilde g^{(d)}(\betavec_j), \tilde h^{(d)}(\betavec_l)) \\
  \cov(\tilde h^{(d)}(\betavec_j), \tilde g^{(d)}(\betavec_l)) & \cov(\tilde h^{(d)}(\betavec_j), \tilde h^{(d)}(\betavec_l))\end{pmatrix}
$$
where,
\begin{align*}
\begin{split}
\cov(\tilde g^{(d)}(\betavec_j), \tilde g^{(d)}(\betavec_l)) &= K^{(d)}(\fvec^{(d)}(\betavec_j), \fvec^{(d)}(\betavec_l)),\\
\cov(\tilde h^{(d)}(\betavec_j), \tilde g^{(d)}(\betavec_l)) &= \frac{\partial K^{(d)}(\fvec^{(d)}(\betavec_j), \fvec^{(d)}(\betavec_l))}{\partial \beta^{(d)}_{j}}, \\
\cov(\tilde g^{(d)}(\betavec_j), \tilde h^{(d)}(\betavec_l)) &= \frac{\partial K^{(d)}(\fvec^{(d)}(\betavec_j), \fvec^{(d)}(\betavec_l))}{\partial \beta^{(d)}_{l}}, \\
\cov(\tilde h^{(d)}(\betavec_j), \tilde h^{(d)}(\betavec_l)) &= \frac{\partial^2 K^{(d)}(\fvec^{(d)}(\betavec_j), \fvec^{(d)}(\betavec_l))}{\partial \beta^{(d)}_{j} \partial \beta^{(d)}_{l}}.
\end{split}
\end{align*}

Our model does not naturally enforce positivity of the predicted variances, however this was not found to be a problem in practice. If this does become an issue, one could use basis functions when constructing the surrogate models, and add constraints to the weights in order to ensure positivity everywhere \citep{azm2019deep}. However, basis function surrogate models tend to be less flexible than (full-rank) Gaussian processes.

\subsection{Predicting the Surrogate Likelihood Elsewhere in the Parameter Space}\label{sec:evaluate_surrogate}

After fitting the surrogate models, we can predict the means and variances of the sufficient statistics for an arbitrary parameter vector $\betavec^{*}$ through Gaussian conditioning. Using the joint model in \eqref{eq:ge_nonstat_gp}, we can obtain predictions of both the means and the variances of the sufficient statistics. However, while we found that the variance information was very important for improving the estimates of the parameters when \emph{fitting} the Gaussian process, we found little benefit in using them to predict the means and the variances, and predicting the variances as the derivative of the mean predictions sufficed. Focusing on the prediction of the means also facilitates comparison between the different surrogate models in Section \ref{sec:gp_surrogate_model}, two of which do not consider the variance of the sufficient statistics.

We consider two approaches for predicting the means and variances of the sufficient statistics with our surrogate models. In the first approach, we let the surrogate mean be equal to the prediction mean of the Gaussian process. That is, for some possibly new parameter value $\betavec^*$, we let
\begin{equation*}
\tilde\mu^{(d)}(\betavec^{*}) = \E[\tilde g^{(d)}(\betavec^{*}) | \Mvec^{(d)}], \quad d = 1, \dots, D,
\end{equation*}
where $\Mvec^{(d)} = \{m^{(d)}(\betavec_1), \dots, m^{(d)}(\betavec_p) \}$. Then, we set the surrogate variance to be equal to the gradient of the surrogate mean, that is, we set
\begin{equation*}
\tilde \sigma^{2 (d)}(\betavec^{*}) = \frac{\partial}{\partial \beta^{(d) *}} \E[\tilde g^{(d)}(\betavec^{*}) | \Mvec^{(d)}], \quad d = 1, \dots, D.
\end{equation*}
Then, the surrogate synthetic likelihood is given by,
\begin{equation*}
\tilde{p}(\zvec \mid \betavec^{*}) = \prod_{d=1}^{D} \mathcal{N}(s^{(d)}(\zvec); \tilde \mu^{(d)}(\betavec^{*}), \tilde \sigma^{2 (d)}(\betavec^{*})).
\end{equation*}

The second approach we propose accounts for the uncertainty in the surrogate model. In this approach, we sample $r$ realizations of the surrogate means, $\{ \hat \mu_1^{(d)}(\betavec^{*}), \dots, \hat \mu_r^{(d)}(\betavec^{*}) \}$, $d = 1, \dots, D$, from the fitted Gaussian process. Then, for each realization $\hat \mu_i^{(d)}(\betavec^*)$, we evaluate the surrogate variance
\begin{equation*}
\hat \sigma_i^{2 (d)}(\betavec^{*}) = \frac{\partial}{\partial \beta^{(d) *}} \hat \mu_i^{(d)}(\betavec^{*}), \quad i = 1,\dots, r.
\end{equation*}
The surrogate synthetic likelihood is then given by,
\begin{equation*}
\tilde{p}_i(\zvec \mid \betavec^{*}) = \prod_{d=1}^{D} \mathcal{N}(s^{(d)}(\zvec); \hat \mu_i^{(d)}(\betavec^{*}), \hat \sigma_i^{2 (d)}(\betavec^{*})),\quad i =1,\dots,r.
\end{equation*}
Because we sample independently from the fitted Gaussian process, the surrogate likelihood for all the realizations is then just the average of the surrogate likelihoods for each realization. Specifically,
\begin{equation*}
\tilde{p}(\zvec \mid \betavec^{*}) \approx \frac{1}{r} \sum_{i=1}^{r} \tilde{p}_i(\zvec \mid \betavec^{*}).
\end{equation*}

\subsection{Inference Using the Surrogate Model}\label{sec:inference_surrogate}

Once we obtain the surrogate synthetic likelihood, there are different ways we can use it when making Bayesian inference. The simplest way is to just substitute the approximately-true likelihood (typically obtained from simulation via the SW algorithm), with the surrogate likelihood in the Metropolis-Hastings ratio in the exchange algorithm, as in \cite{moores2020scalable} and \cite{park2020function}. This is computationally efficient, as it precludes the need for further simulations of the sufficient statistics, but it is an inexact-approximate method. One exact-approximate MCMC method, which uses the surrogate likelihood to reduce the computational cost, is delayed-acceptance MCMC \citep{christen2005markov}. The delayed-acceptance algorithm is a two-stage algorithm, wherein the first step involves using the surrogate likelihood to quickly reject any poor proposals: Only good proposals accepted in the first stage are passed through to the second stage, where similarly to the exchange algorithm, auxiliary variables are used to accept or reject the proposals. This algorithm prevents one from wasting computational resources on simulations of sufficient statistics at a poor proposal. The delayed-acceptance algorithm is detailed in Algorithm \ref{alg:delayed_accept}.

\begin{algorithm}
Denote samples from the target posterior distribution as $\tilde\betavec_i: i = 0,\dots,T$, where $T$ is the number of iterations. Initialize $\tilde\betavec_0$. For $i = 0, \dots, (T - 1)$ do \\
\nl Propose a new value $\betavec^{\new}$ from a proposal distribution $q(\betavec^{\new} \mid \tilde\betavec_i)$. \\
\nl Stage 1: Pass $\betavec^{\new}$ to Stage 2 with probability
\[ \min \left(1, \frac{q(\tilde\betavec_i \mid \betavec^{\new}) p(\betavec^{\new}) \tilde{p}(\zvec \mid \betavec^{\new})}{q(\betavec^{\new} \mid \tilde\betavec_i) p(\tilde\betavec_i) \tilde{p}(\zvec \mid \tilde\betavec_i)} \right). \]
\nl Stage 2: Simulate pseudo-data $\xvec^{\new}$ from the likelihood $p(\xvec^{\new} \mid \betavec^{\new})$. Accept $\betavec^{\new}$ with probability
\[ \min \left(1, \frac{\tilde{p}(\zvec \mid \tilde\betavec_i) \exp\{\betavec^{\new^T} \svec(\zvec)\} \exp\{\tilde\betavec_i^T \svec(\xvec^{\new}) \} }{\tilde{p}(\zvec \mid \betavec^{\new}) \exp\{\tilde\betavec_i^T \svec(\zvec)\} \exp\{\betavec^{\new^T} \svec(\xvec^{\new})\} } \right). \]
\nl If $\betavec^{\new}$ is accepted at Stage 2, then set $\tilde\betavec_{i+1} = \betavec^{\new}$, otherwise, if $\betavec^{\new}$ is not accepted at either Stage 1 or Stage 2, then set $\tilde\betavec_{i+1} = \tilde\betavec_i$.

\caption{{\bf Delayed-acceptance MCMC} \label{alg:delayed_accept}}
\end{algorithm}

An alternative to MCMC methods is importance sampling \citep[e.g.,][]{everitt2017bayesian}. Importance sampling is an exact-approximate method, yet it affords a significant improvement in computational time, as the simulations of sufficient statistics (required to determine the importance weights), can be performed in parallel. The importance sampling algorithm shown in Algorithm \ref{alg:importance} uses auxiliary variables in a similar fashion to \cite{moller2006efficient}. In our approach, we choose the proposal distribution for importance sampling to be the \emph{surrogate posterior distribution}, which we define to be that posterior distribution obtained when using the surrogate likelihood function directly in place of the true likelihood function. A fast approach to obtain the surrogate posterior is by using grid approximation, wherein one evaluates the surrogate posterior density at regular points on a grid \citep[e.g.,][]{gelman2013bayesian}; this is feasible in the low-dimensional settings we consider. Once the grid approximation is found, it is straightforward to draw samples from the surrogate posterior distribution prior to reweighing using the simulations  of the sufficient statistics. 


\begin{algorithm}
Denote samples from the target posterior distribution as $\tilde\betavec_i: i = 1,\dots,T$, where $T$ is the number of samples.

\nl Sample $\tilde\betavec_1, \dots, \tilde\betavec_T$ from the surrogate posterior distribution $\tilde{p}(\zvec \mid \betavec) p(\betavec) $. \\
\nl For each $\tilde\betavec_i$, simulate pseudo-data $\xvec_i$ from the likelihood $p(\xvec_i \mid \tilde\betavec_i)$. \\
\nl Calculate the importance weight for each of the samples,
\[ w_i = \frac{\exp\{\tilde\betavec_i^T \svec(\zvec)\} \exp\{\hat{\betavec}^T \svec(\xvec_i)\}  }{ \exp\{\tilde\betavec_i^T \svec(\xvec_i)\} \tilde{p}(\zvec \mid \tilde\betavec_i) }, \]
where $\hat{\betavec}$ is fixed at an arbitrary point. Similar to \citet{moller2006efficient}, we choose the maximum likelihood estimate, which we approximate using the surrogate likelihood. \\
\nl Normalize the weights
\[ \tilde w_i = \frac{w_i}{\sum_{i=1}^{T} w_i}. \]
\nl Obtain the weighted posterior samples from which we can then make inference on $\betavec$.

\caption{{\bf Importance sampling} \label{alg:importance}}
\end{algorithm}

\section{Data Examples}\label{sec:examples}

In this section, we show examples that demonstrate the utility of our proposed Gaussian process surrogate models when making inference on parameters that appear in  the Potts, the hidden Potts, and the autologistic models, using  both simulated and real-world data sets. 
All MCMC computations were done on a computer with a 6-core Intel Core i7-8700 @3.2 GHz, 32 GB RAM, and an NVIDIA GeForce GTX 1600 GPU. All computations with the importance sampling algorithm were done on a server with 64 cores in Intel Xeon E5-2683 @2.1 GHz processors, 256 GB RAM, and an NVIDIA GeForce GTX TITAN GPU, in order to take advantage of algorithm parallelization.
Code to reproduce the results in this section can be found at \url{
https://github.com/quanvu17/warped_gradient_enhanced_GP_surrogate}.

\subsection{Potts Model}\label{sec:potts}

The Potts model is often used for analyzing spatial dependence between neighboring labels in images. In the Potts model, the probability of observing a specific combination of labels is defined as
\[ p(\zvec \mid \beta^{(1)}) = \frac{\exp(\beta^{(1)} \sum_{u \sim v} \delta(z_u, z_v) )}{\mathcal{C}(\beta^{(1)})}, \]
where $z_u, u = 1, \dots, N$, is the label of pixel $u$, $u \sim v$ denotes the neighboring pixels of pixel $u$, and $\delta(\cdot)$ is the Kronecker delta function. The sufficient statistic of the Potts model, $s(\zvec) = \sum_{u \sim v} \delta(z_u, z_v)$, is the count of pairs of neighboring pixels that have the same label. The normalizing constant $\mathcal{C}(\beta^{(1)}) = \sum_{\zvec \in \mathcal{Z}} \exp(\beta^{(1)} \sum_{u \sim v} \delta(z_u, z_v) )$ involves a summation over all possible combinations of the labels over all the pixels, and therefore is computationally infeasible to evaluate. The Potts model undergoes a phase transition from a disordered state (where most neighboring pixels do not share the same label) to an ordered state (where most neighboring pixels share the same label) near a critical value of the parameter $\beta^{(1)}$, which is often referred to as the inverse temperature parameter. As $p(\zvec \mid \beta^{(1)})$ involves a computationally intractable sum and also contains a phase transition, we use the proposed gradient-enhanced nonstationary Gaussian process surrogate model to approximate this probability.

In this experiment we consider the Potts model for a $1000 \times 1000$ size image, where the number of labels is $k = 5$. The likelihood function exhibits a phase transition around $\beta^{(1)} = 1.175$; see Figure~\ref{fig:potts_summary_stats}.

\subsubsection{Comparison of Gaussian Process Surrogate Models}\label{sec:potts_comparison}

We chose a bounded uniform prior distribution on the interval $[0.9, 1.3]$ for the inverse temperature parameter $\beta^{(1)}$ (see Section~\ref{sec:priors} for a discussion on the choice of prior distributions). We chose $p=51$ equally-spaced points on this interval for training, and 50 points (in between the training data) for testing. For each of the 101 values of $\beta^{(1)}$ we simulated sufficient statistics using the SW algorithm.

We fit the surrogate models introduced in Section \ref{sec:gp_surrogate_model} to the training data set:
\begin{enumerate}
\item A stationary Gaussian process model (S-GP),
\item A nonstationary Gaussian process model (NS-GP),
\item A gradient-enhanced nonstationary Gaussian process model (GE-NS-GP),
\end{enumerate}
where the trend $b^{(1)}(\cdot)$ was fixed to be the mean of the observed sample means, and the stationary Gaussian process $g^{(1)}(\cdot)$ was set to have the Mat{\'e}rn 3/2 covariance function
\begin{equation*}
K^{(1)}(\betavec_j, \betavec_l) = \cov(g^{(1)}(\betavec_j), g^{(1)}(\betavec_l)) = \xi^{2 (1)} (1 + a^{(1)} \norm{\hvec}) \exp(-a^{(1)} \norm{\hvec}),
\end{equation*}
where $\xi^{2 (1)}$ is the process variance parameter, $a^{(1)}$ is the process scale parameter, and $\hvec \equiv \betavec_l - \betavec_j$. For Models 2 and 3 we used an axial warping unit for $f^{(1)}(\cdot)$ \citep{azm2019deep} comprised of 100 sigmoid basis functions. 

For comparison purposes we also ran the parametric functional approximate Bayesian (PFAB) algorithm \citep{moores2020scalable}, which uses a surrogate parametric model that takes into account specific properties of the Potts likelihood (e.g., the critical value, which can be calculated exactly for a 2D lattice). PFAB is specifically designed for the Potts model and is therefore a good candidate for comparison. Note that our Gaussian surrogate models are more general as they can be easily used with other exponential-family models (e.g., with the autologistic model and the Kent distribution, as we show in Section~\ref{sec:autologistic} and Appendix \ref{sec:Kent}, respectively).

\begin{figure}[t!]
	\centering
	
	\includegraphics[width=0.4\textwidth]{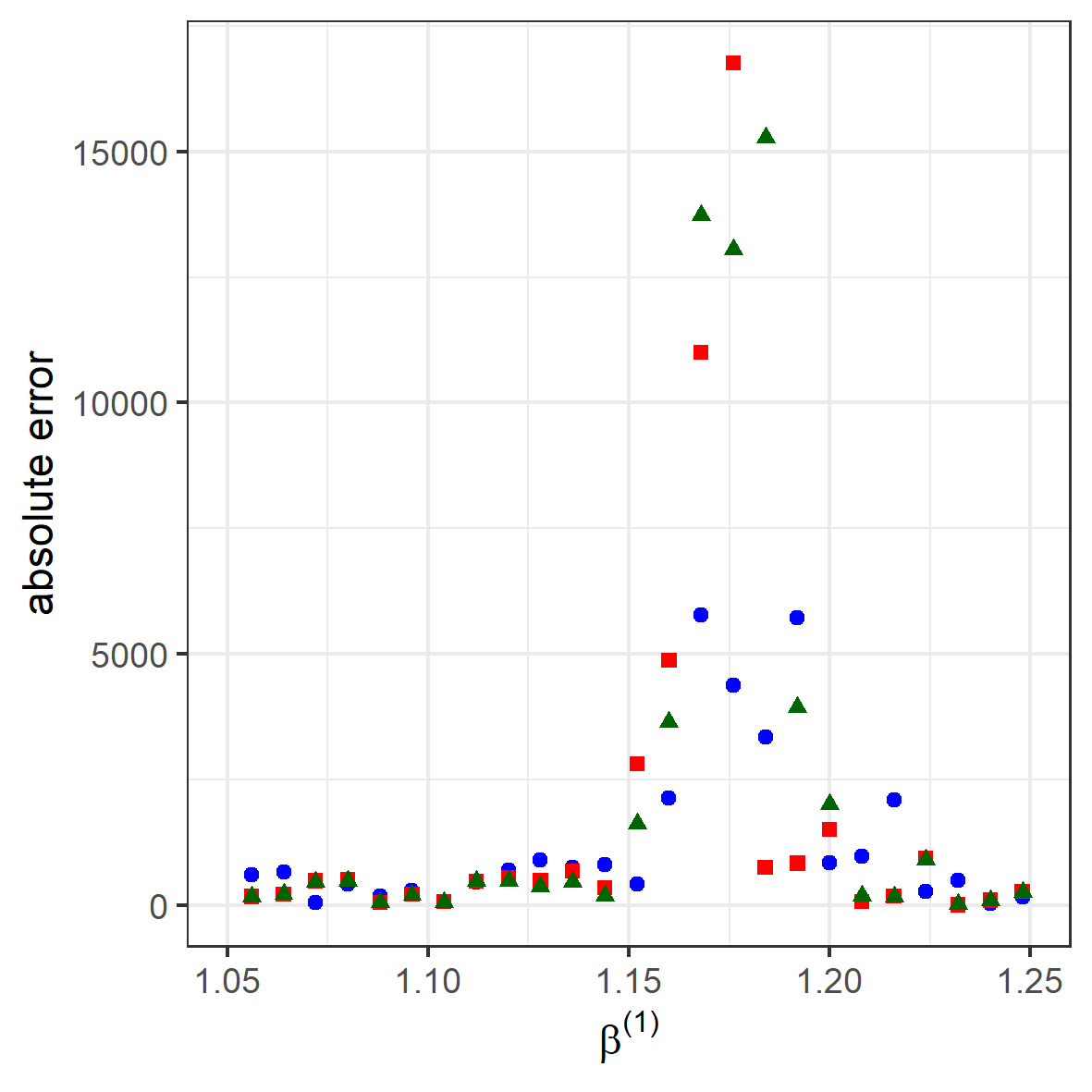}
	\includegraphics[width=0.4\textwidth]{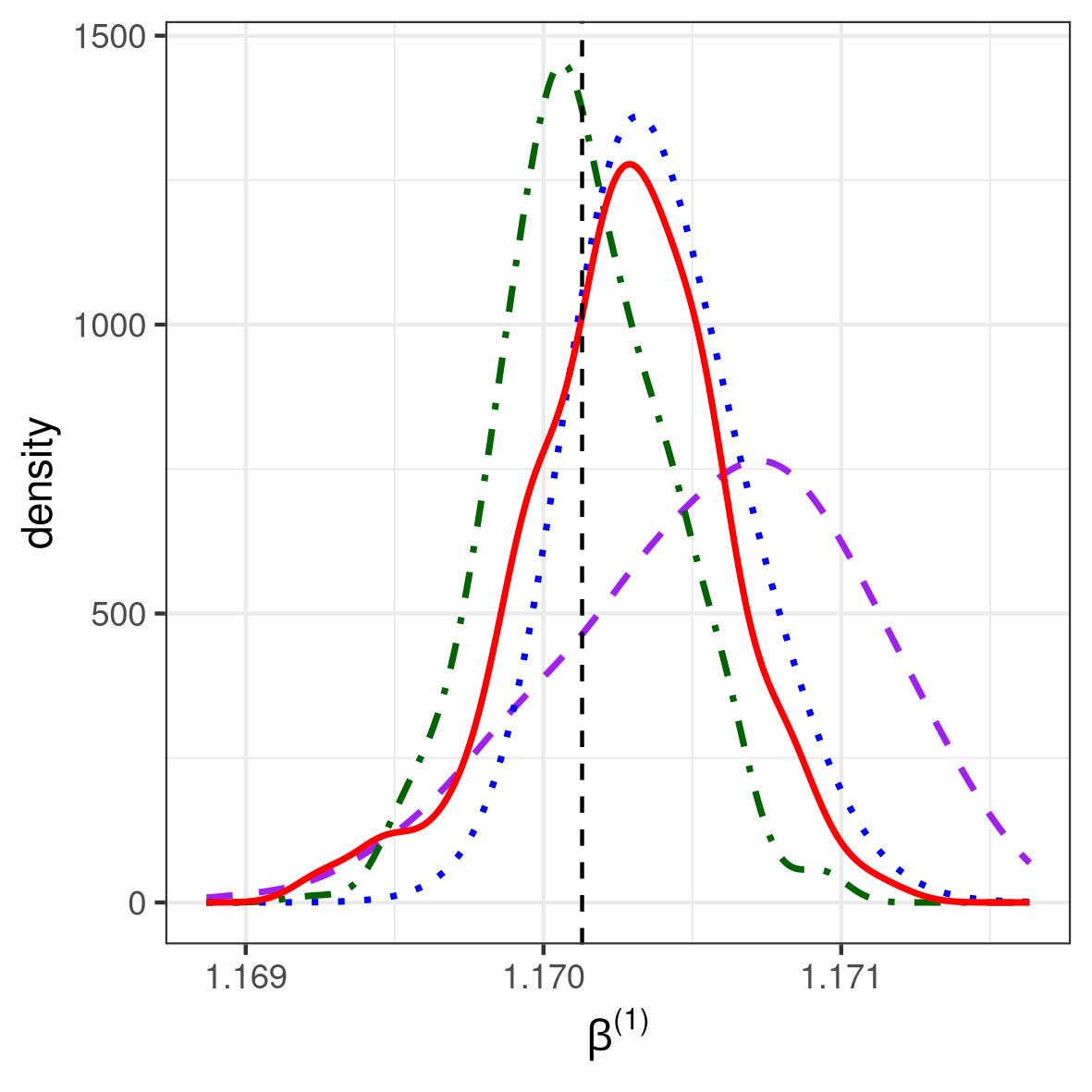}

	\caption{
	Left panel: Absolute errors of the means of the sufficient statistics for the surrogate models. Green triangle: S-GP. Red square: NS-GP. Blue dot: GE-NS-GP. Right panel: Posterior distribution of the parameter $\beta^{(1)}$ for the Potts model. Purple dashed line: GE-NS-GP surrogate posterior. Blue dotted line: Importance sampling. Green dot-dashed line: Delayed-acceptance MCMC. Red solid line: Exchange algorithm.  Vertical dashed line: True parameter value.
	}
	\label{fig:potts}
\end{figure}

Fitting of the Gaussian process surrogate models 1--3 only took a few seconds using maximum likelihood, while running the PFAB algorithm took nearly 30 minutes. The predicted means and variances of the sufficient statistics at the testing locations were then compared to those of the simulated sufficient statistics at these testing locations. The mean absolute prediction error (MAPE) and root mean square prediction error (RMSPE) are shown in Table \ref{tbl:potts_compare}. The PFAB algorithm produced substantially worse predictions than all of the Gaussian process models for both the mean and the variance, despite it being specifically designed for the Potts model. Focusing just on the Gaussian process models, we see that in terms of predicting the mean, the stationary GP performed the worst, while the gradient-enhanced nonstationary GP model performed the best. The left panel of Figure \ref{fig:potts} shows the absolute errors for Models 1--3. All three models perform  similarly in regions far away from the phase transition. However, near the phase transition, we see that the stationary model results in large errors, while the gradient-enhanced nonstationary model results in smaller errors.
Table \ref{tbl:potts_compare} shows that the GE-NS-GP model also generates slightly better predictions of the variances, when compared to those of the S-GP and NS-GP models, although the improvement is less substantial. 

\begin{table}[t!]
	\centering
	\caption{Comparison of the different surrogate models.}
	\label{tbl:potts_compare}
    \bgroup
    \def\arraystretch{1}
	\begin{tabular}{ |c|c|c|c|c| }
		\hline
		Model & PFAB & S-GP & NS-GP & GE-NS-GP \\
		\hline
		MAE (mean, $\times 10^3$) & 4.224 & 1.258 & 1.007 & 0.817 \\
		\hline
		RMSPE (mean, $\times 10^3$) & 12.581 & 3.554 & 2.978 & 1.528 \\
		\hline
		MAE (variance, $\times 10^6$) & 6.387 & 5.810 & 5.783 & 5.699 \\
		\hline
		RMSPE (variance, $\times 10^7$) & 4.185 & 3.819 & 3.792 & 3.748 \\
		\hline
	\end{tabular}
    \egroup
\end{table}

\subsubsection{Inference}

In this sub-section, we conduct experiments where we infer the inverse temperature parameter from images using the Gaussian process surrogate models.
We set the true inverse temperature parameter $\beta^{(1)}$ by simulating from a normal distribution centered at the true critical value $\beta^{(1)} = 1.1744$ \citep[see][]{potts1952}, with standard deviation 0.05: The simulated value was 1.1701. We then generated five different images from this parameter value to show the repeatability of our experiment.

We used the GE-NS-GP surrogate model to do Bayesian inference using importance sampling and delayed-acceptance MCMC, and compared both these methods to the exchange algorithm. We first generated 1000 samples from the surrogate posterior distribution using grid approximation, which took less than one second. Then, for importance sampling, we simulated pseudo-data to reweigh the samples. For both the delayed-acceptance MCMC algorithm and the exchange algorithm, we used 2200 MCMC iterations (including 200 burn-in iterations). The posterior distributions obtained from these methods from one of the five images are shown in the right panel of Figure \ref{fig:potts} and summarized in Table \ref{tbl:potts_posterior}. Posterior distributions for the other four images are shown in Figure \ref{fig:potts_2} in Appendix \ref{sec:appendix}. We see that running the importance sampler only took 0.7 hours, since it is parallelizable. The sampler also resulted in a much higher effective sample size (ESS) per hour (79.8 samples per hour) than the other exact-approximate methods. Delayed-acceptance took 25.0 hours, a bit less, relatively, than the 31.0 hours of the exchange algorithm, and resulted in a small gain in ESS per hour of 7.2, when compared to 5.7 for the exchange algorithm. In Figure \ref{fig:potts}, we also show the inexact-approximate posterior distribution (i.e., the surrogate posterior distribution), where we simply use our surrogate model as the likelihood and do grid-based inference. Note that this inexact posterior distribution is slightly different from the posterior distributions obtained from the exact methods.
However, the speed at which this approximate posterior distribution was obtained ($< 1$ second in this case) can make it attractive for situations where computing time is a serious concern.

\begin{table}
	\centering
	\caption{Posterior distribution for the Potts example of Section \ref{sec:potts}.}
	\label{tbl:potts_posterior}
	\bgroup
	\def\arraystretch{1}
	\begin{tabular}{ |c|c|c|c|c| }
		\hline
		Method & Surrogate & Importance sampling & Delayed-acceptance & Exchange \\
		\hline
		Posterior mean & 1.17056 & 1.17039 & 1.17013 & 1.17027 \\
		\hline
		Posterior SD & 0.000497 & 0.000252 & 0.000285 & 0.000333 \\
		\hline
		Time (hours) & $3 \times 10^{-6}$ & 0.7 & 25.0 & 31.0 \\
		\hline
		ESS/hour & $3.6 \times 10^9$ & 79.8 & 7.2 & 5.7 \\
		\hline
	\end{tabular}
	\egroup
\end{table}

\subsection{Hidden Potts Model}\label{sec:hidden_potts}

\begin{figure}[t!]
	\centering
	
	\includegraphics[width=1\textwidth]{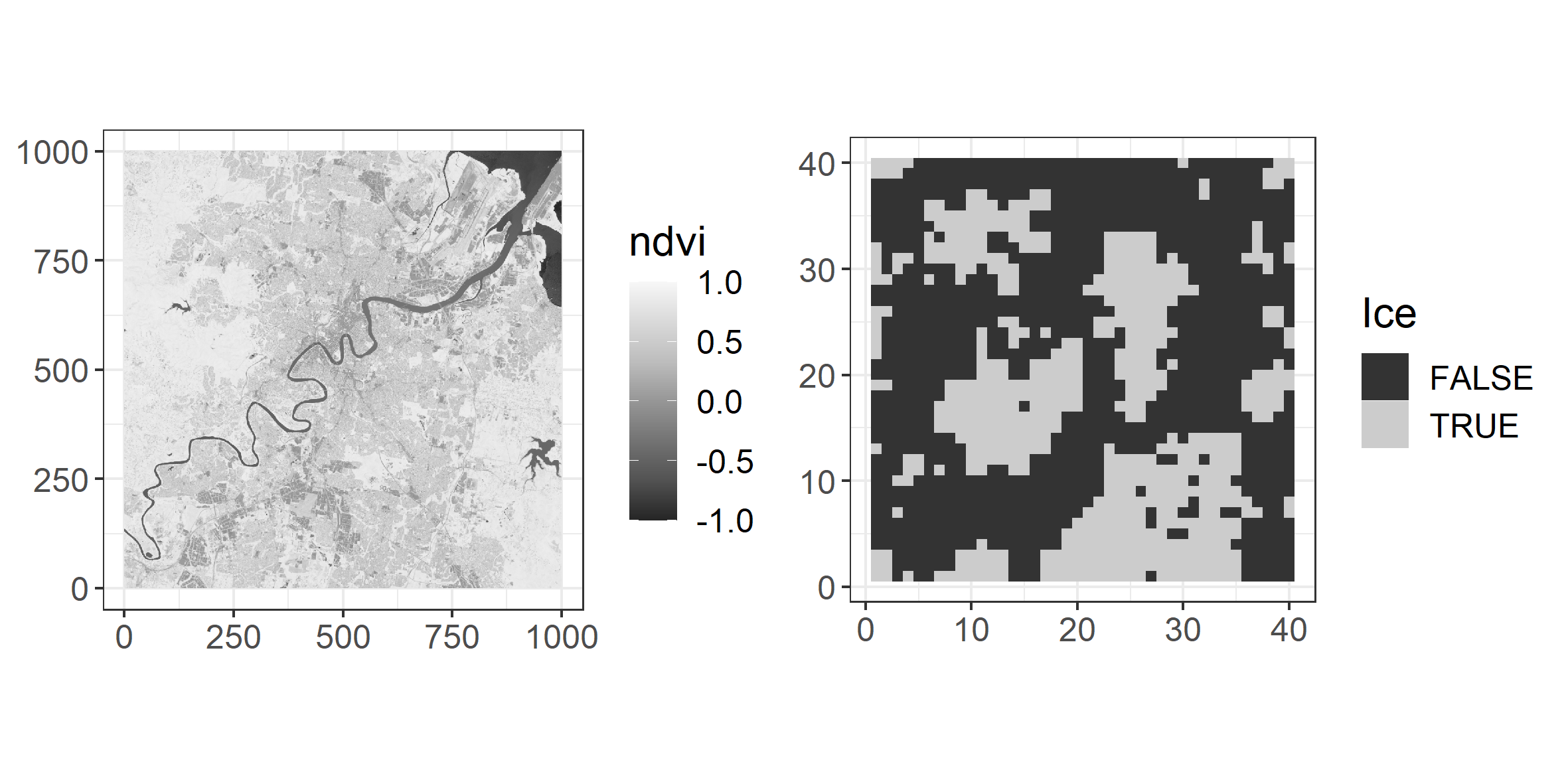}

	\caption{
	  Images used in the analysis in Section \ref{sec:hidden_potts} and Section \ref{sec:autologistic}, respectively. Left panel: NDVI image of Brisbane derived from Landsat-8 data on 03 May 2015. Right panel: Cropped satellite image of ice floe, originally published in \cite{banfield1992ice}, available in the \texttt{R} package \texttt{PAWL}.
	}
	\label{fig:data_plots}
\end{figure}

The hidden Potts model is an extension of the Potts model, and is used for analyzing  spatial dependence when the labels are not directly observed. The model links the observed pixel intensity, $y_u$, with the latent label, $z_u$, through the following relationship
\[ p(y_u \mid z_u = \lambda, \mu_{\lambda}, \sigma^2_{\lambda}) = \Gau(\mu_{\lambda}, \sigma^2_{\lambda} ),~~ \lambda = 1,\dots,k, \] 
\noindent where the $\{\mu_\lambda\}_\lambda$ and the $\{\sigma^2_\lambda\}_\lambda$ are unknown and equipped with informative prior distributions; see \citet{moores2020scalable} for details. In this example, we consider a normalized difference vegetation index (NDVI) $1000 \times 1000$ image of Brisbane derived from Landsat-8 satellite data on 03 May 2015\footnote{Data available from \url{https://hpc.niasra.uow.edu.au/ckan/dataset/ndvip089r079_20150503}}, shown in the left panel of Figure \ref{fig:data_plots}. We analyzed spatial dependence in the data by classifying pixels using five labels: forest, light vegetation, urban area, suburban area, and water. We use the hidden Potts model and make inference on the inverse temperature parameter $\beta^{(1)}$, which determines the spatial dependence between these labels. As we do not directly observe the labels we cannot use importance sampling for this model. We therefore used MCMC to update the latent labels, as well as the parameters  $\{\mu_\lambda\}_\lambda$ and $\{\sigma^2_\lambda\}_\lambda$, at each iteration of the chain.  

\begin{figure}[t!]
	\centering
	
	\includegraphics[width=0.4\textwidth]{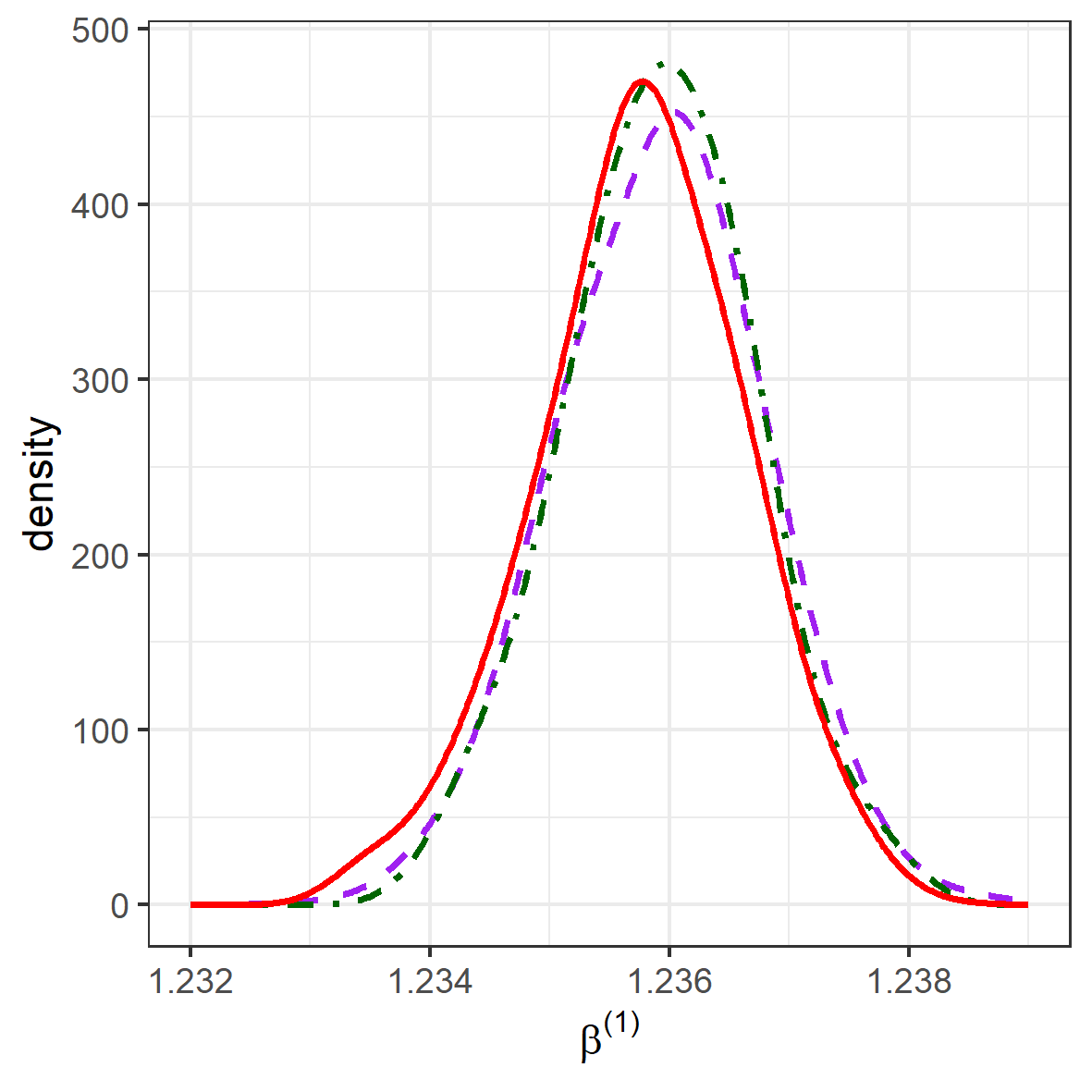}

	\caption{
	Posterior distribution of the parameter $\beta^{(1)}$ for the hidden Potts model. Purple dashed line: Surrogate posterior. Green dot-dashed line: Delayed-acceptance MCMC. Red solid line: Exchange algorithm.
	}
	\label{fig:hidden_potts_posterior}
\end{figure}

We chose the prior distribution over $\beta^{(1)}$ to be a uniform distribution on the interval $[0.9, 1.3]$. The satellite image in this section has the same dimension and number of labels as the simulated image in Section \ref{sec:potts}; hence we used the same GE-NS-GP model we fitted in Section \ref{sec:potts_comparison} when making inference. We compared the inexact-approximate method to the two MCMC exact-approximate methods we consider in this work: delayed-acceptance MCMC and  the exchange algorithm. We ran 2200 MCMC iterations (including 200 burn-in iterations) for each method. The posterior distributions obtained from these methods are shown in Figure \ref{fig:hidden_potts_posterior} and summarized in Table \ref{tbl:hidden_potts_posterior}. We see that all the posterior distributions are very similar. However, with the same number of iterations, evaluating the surrogate inexact-approximate posterior distribution took 3.2 hours. On the other hand, delayed-acceptance MCMC took 14.6 hours, while the exchange algorithm took 28.3 hours. Unsurprisingly, the highest ESS per hour of 112.8 was obtained with the inexact-approximate method. Delayed acceptance and the exchange algorithm yielded an ESS per hour of  7.8 and 6.5, respectively.

\begin{table}
	\centering
	\caption{Posterior distribution for the hidden Potts example of Section~\ref{sec:hidden_potts}.}
	\label{tbl:hidden_potts_posterior}
    \bgroup
    \def\arraystretch{1}
	\begin{tabular}{ |c|c|c|c| }
		\hline
		Method & Surrogate & Delayed-acceptance & Exchange \\
		\hline
		Posterior mean & 1.23593 & 1.23592 & 1.23576 \\
		\hline
		Posterior SD & 0.000842 & 0.000777 & 0.000845 \\
		\hline
		Time (hours) & 3.2 & 14.6 & 28.3 \\
		\hline
		ESS/hour & 112.8 & 7.8 & 6.5 \\
		\hline
	\end{tabular}
    \egroup
\end{table}

\subsection{Autologistic Model}\label{sec:autologistic}

The autologistic model is another extension of the Potts model, where the sufficient statistics also include the number of pixels associated with each label. The two-label autologistic model, proposed by \cite{besag1974spatial}, has as likelihood function
\[ p(\zvec \mid \betavec) = \frac{\exp(\beta^{(1)} \sum_{u} z_u + \beta^{(2)} \sum_{u \sim v} \delta(z_u, z_v) )}{\mathcal{C}(\betavec)}, \]
where the label $z_u$ of pixel $u$ takes the value $1$ or $-1$.

We use the autologistic model to analyze spatial dependence in a satellite image of ice floe originally published in \cite{banfield1992ice}; here we consider the cropped $40 \times 40$ image \citep{bornn2013adaptive}, available in the \texttt{R} package \texttt{PAWL}, and shown in the right panel of Figure \ref{fig:data_plots}.

We chose the prior distribution to be a uniform distribution on $[-0.2, 0.1] \times [0.7, 1.2]$, and we chose training data from a $7 \times 11$ equally-spaced grid on this rectangular domain to fit the GE-NS-GP surrogate model. We fixed the trend $b^{(1)}(\cdot)$, $b^{(2)}(\cdot)$ to the means of the observed sample means for each sufficient statistic, and chose the covariance functions for $g^{(1)}(\cdot)$ and $g^{(2)}(\cdot)$ to be Mat{\'e}rn 3/2 covariance functions, and used axial warping units in each dimension for the the deformation functions $\fvec^{(1)}(\cdot)$ and $\fvec^{(2)}(\cdot)$.

\begin{figure}[t!]
	\centering
	
	\includegraphics[width=0.8\textwidth]{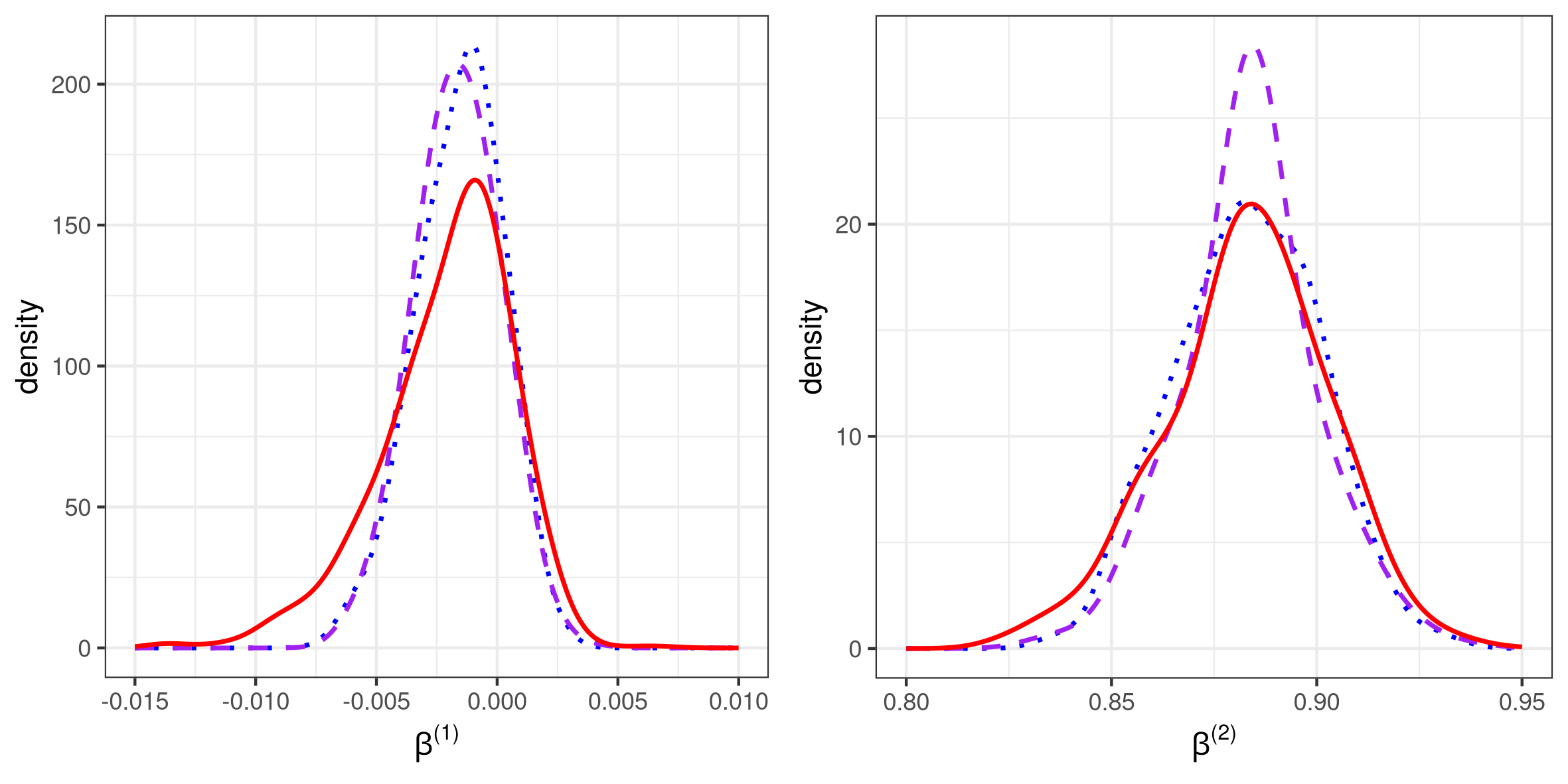}

	\caption{
	Posterior distribution of the parameter $\beta^{(1)}$ (left panel) and $\beta^{(2)}$ (right panel) for the autologistic model. Purple dashed line: Surrogate posterior. Blue dotted line: Importance sampling. Red solid line: Exchange algorithm.
	}
	\label{fig:autologistic_posterior}
\end{figure}

We first took 2000 samples from the surrogate posterior distribution using grid approximation. For importance sampling, we simulated sufficient statistics at these 2000 values to reweigh the samples. We compared the results to those using the exchange algorithm, which was run for 4200 iterations (200 were discarded as burn-in). The posterior distributions obtained from the two methods are shown in Figure \ref{fig:autologistic_posterior} and summarized in Table \ref{tbl:autologistic_posterior}. As we saw in Section \ref{sec:potts}, all posterior distributions are largely similar, and we again see the huge computational gain of importance sampling  over the exchange algorithm in terms of time taken and ESS per unit time. 

\begin{table}[t!]
	\centering
	\caption{Posterior distribution for the autologistic example of Section \ref{sec:autologistic}.}
	\label{tbl:autologistic_posterior}
	\bgroup
	\def\arraystretch{1}
	\begin{tabular}{ |c|c|c|c| }
		\hline
		Method & Surrogate & Importance sampling & Exchange \\
		\hline
		Posterior mean $\beta^{(1)}$ & -0.00167 & -0.00154 & -0.00208 \\
		\hline
		Posterior SD $\beta^{(1)}$ & 0.00176 & 0.00176 & 0.00272 \\
		\hline
		Posterior mean $\beta^{(2)}$ & 0.88361 & 0.88312 & 0.88315 \\
		\hline
		Posterior SD $\beta^{(2)}$ & 0.01647 & 0.01748 & 0.02028 \\
		\hline
		Time (mins) & 0.001 & 1.8 & 29.2 \\
		\hline
		ESS/minute $\beta^{(1)}$ & $3 \times 10^6$ & 201 & 7.1 \\
		\hline
		ESS/minute $\beta^{(2)}$ & $3 \times 10^6$ & 201 & 5.3 \\
		\hline
	\end{tabular}
	\egroup
\end{table}

\section{Discussion}

In this article, we have introduced a warped gradient-enhanced Gaussian process surrogate model for modeling the means and variances of the sufficient statistics of models with intractable likelihoods. In particular, we showed that the inclusion of nonstationarity and gradient information in this surrogate model resulted in smaller errors at the phase transition than the stationary Gaussian process surrogate model, and that the surrogate model can easily be used to speed up parameter inference, either in inexact-approximate methods, or in exact-approximate methods such as importance sampling and delayed-acceptance MCMC.

In Section \ref{sec:examples}, we showed examples for the Potts, hidden Potts, and autologistic models. Inference for other models with exponential-family intractable likelihoods can also be done using our approach. One such model is the exponential random graph model (ERGM). ERGMs have, however, some distinct properties. In particular, there are some ERGMs for which the sufficient statistics do not result in any phase transitions; inference with these models can be done effectively using stationary Gaussian process surrogate models \citep{park2020function}. Sufficient statistics in other ERGMs experience steep phase transitions, and simulations at the phase transitions often result in bimodal distributions. Inference with these types of models is a future direction of research.

In this work, we have assumed independence between the sufficient statistics (see \eqref{eq:syn_like}). Although this assumption does not seem to have affected our inferences, one can envisage the use of surrogate models where the sufficient statistics are modeled as dependent. One would need a Gaussian process which jointly models all the surrogate means and covariances of the sufficient statistics; in this case, the cross-covariances between the means of the different sufficient statistics are non-zero. The cross-covariances between the variances of the different sufficient statistics are also non-zero.
This leads to the question of what is the appropriate form of the cross-covariance function. The answer to this question is non-trivial, as the covariance between two different sufficient statistics can be written as two different gradients of the means of these sufficient statistics with respect to a different dimension of the parameter, that is,
\begin{align*}
\cov_{\betavec}(s^{(d_1)}(\zvec), s^{(d_2)}(\zvec)) &= \frac{\partial^2}{\partial \beta^{(d_1)} \partial \beta^{(d_2)}} \log \mathcal{C}(\betavec) \\
&= \frac{\partial}{\partial \beta^{(d_1)}} \mu^{(d_2)}(\betavec) = \frac{\partial}{\partial \beta^{(d_2)}} \mu^{(d_1)}(\betavec).
\end{align*}

In future work we will also address the scalability of our approach to high-dimensional parameter space; in this article, we only showed examples with one-dimensional and two-dimensional parameter spaces. In a high-dimensional parameter space, we would need to simulate sufficient statistics at a much larger number of parameter values in order to get a good fit to the means and variances of the sufficient statistics. This will entail investigating the use of approximation methods to Gaussian processes \citep[e.g., ][]{Quinonero_2005} in the context of surrogate likelihood models. 

The relationship between the mean and the variance shown in \eqref{eq:derivative_property} holds only for exponential-family models, and therefore, the gradient-enhanced Gaussian process surrogates is only applicable to models in that class. However, there still are elements in our work that can be used to construct representative synthetic likelihood functions for more general (non-exponential) intractable models. In these cases, we would approximate the likelihood as a multivariate normal distribution of summary statistics. For the means of the summary statistics, we could then use the (warped) nonstationary Gaussian process surrogate model \eqref{eq:nonstat_gp} introduced in Section \ref{sec:gp_surrogate_model}. For the (log) variances, we could also use a nonstationary Gaussian process surrogate model; that is, we could employ the model
	\begin{equation}\label{eq:logvar_gp}
		\log \sigma^{2 (d)}(\cdot) = c^{(d)}(\cdot) + h^{(d)}(\fvec^{(d)}(\cdot)),
	\end{equation}
where $c^{(d)}(\cdot)$ is a trend, $h^{(d)}(\cdot)$ is a zero-mean stationary Gaussian process, and $\fvec^{(d)}(\cdot)$ is a warping function. One would need to use a different warping function for the variance than from the mean, since unlike in the exponential-family case there is no pre-defined relationship between the mean and the variance. The resulting model can be seen as a warped version of the coupled mean and variance Gaussian processes presented by \citet[][Chap.~10]{gramacy2020surrogates}. Developing a flexible class of models on these lines for non-exponential family models is the subject of future research.


\section*{Acknowledgments}

Quan Vu was supported by a University Postgraduate Award from the University of Wollongong, Australia. Andrew Zammit-Mangion's research was supported by an Australian Research Council (ARC) Discovery Early Career Research Award, DE180100203.

\bibliographystyle{apalike}
\bibliography{biblio}

\newpage

\appendix

\section{Additional Figures}\label{sec:appendix}

\begin{figure}[h!]
	\centering
	
	\includegraphics[width=0.4\textwidth]{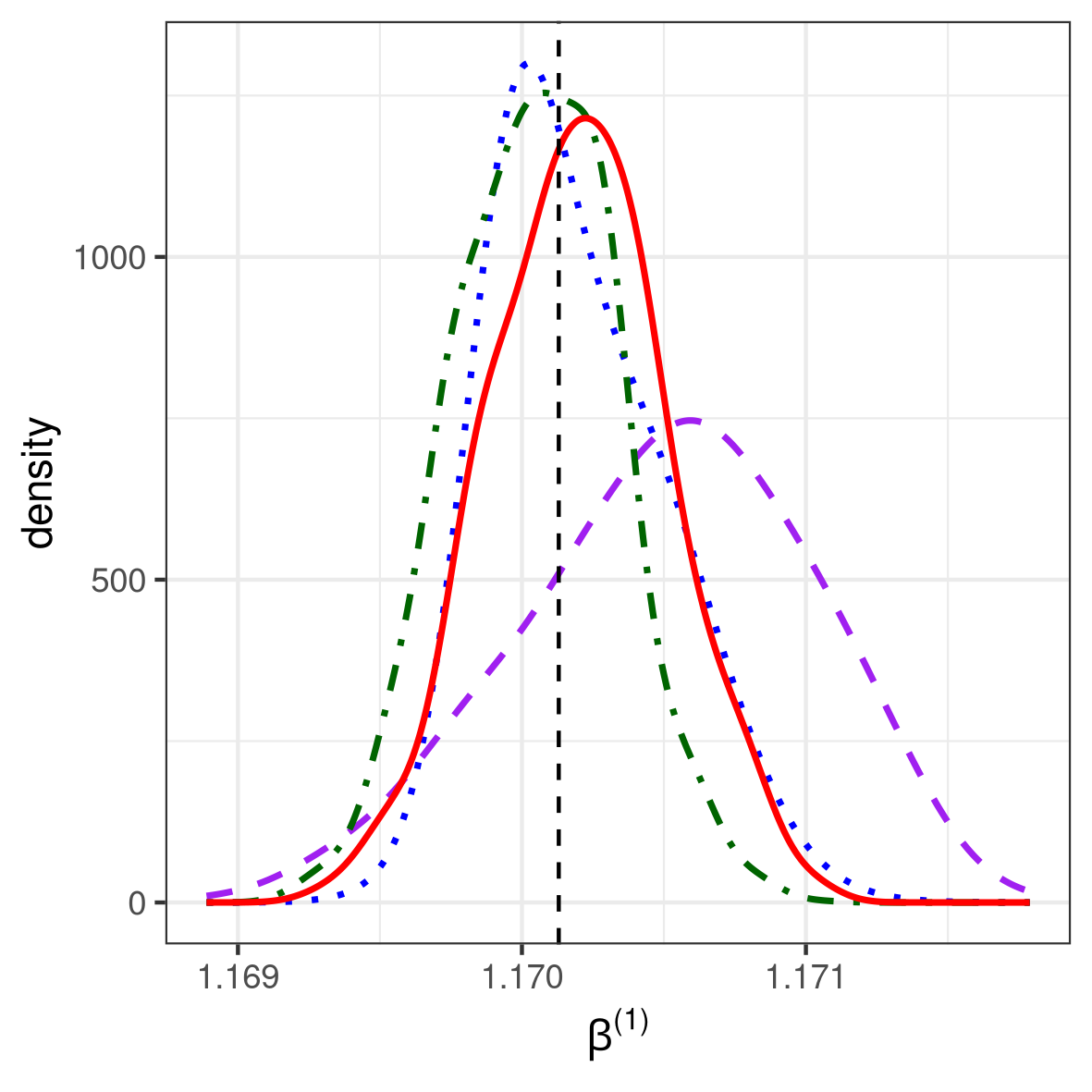}
	\includegraphics[width=0.4\textwidth]{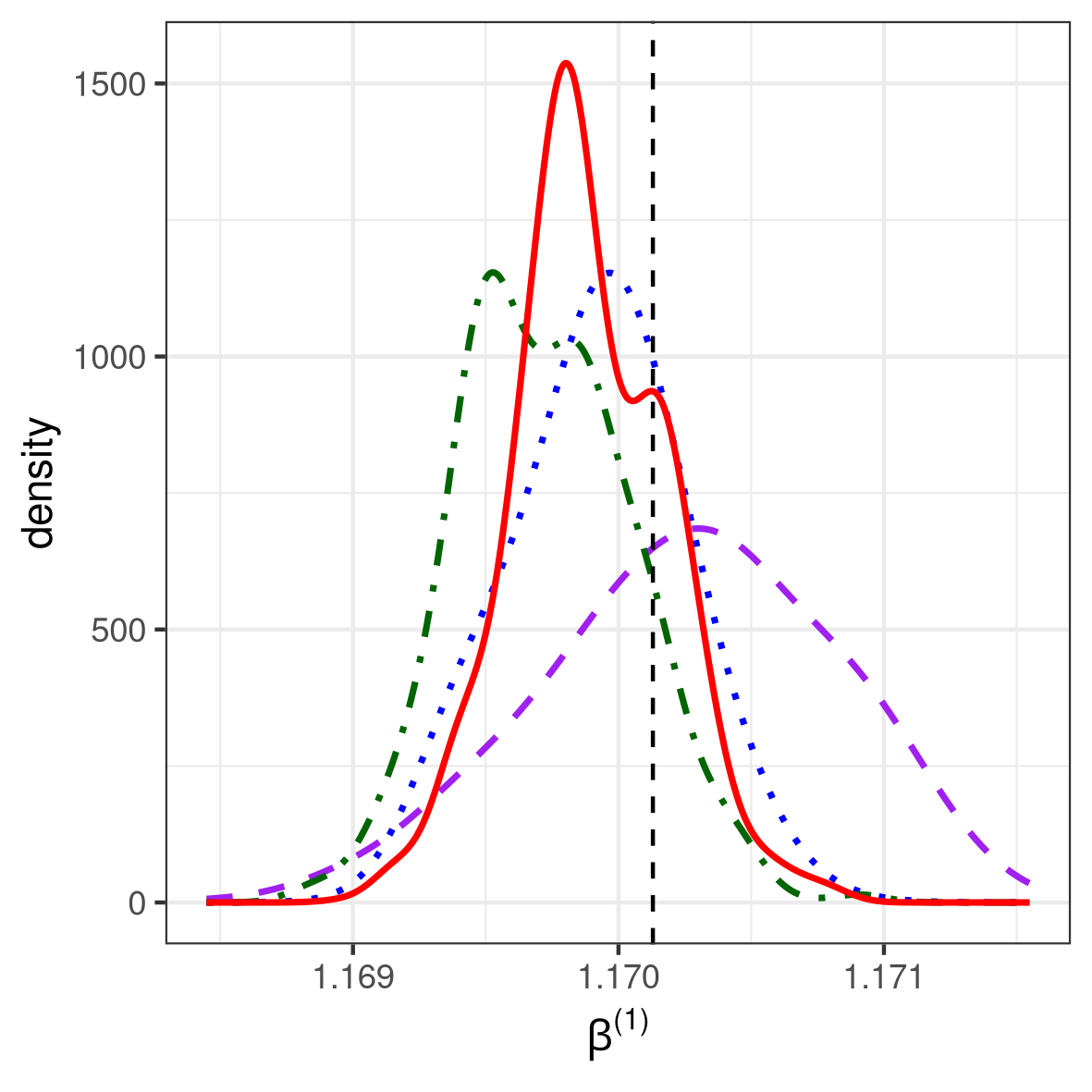}
	\includegraphics[width=0.4\textwidth]{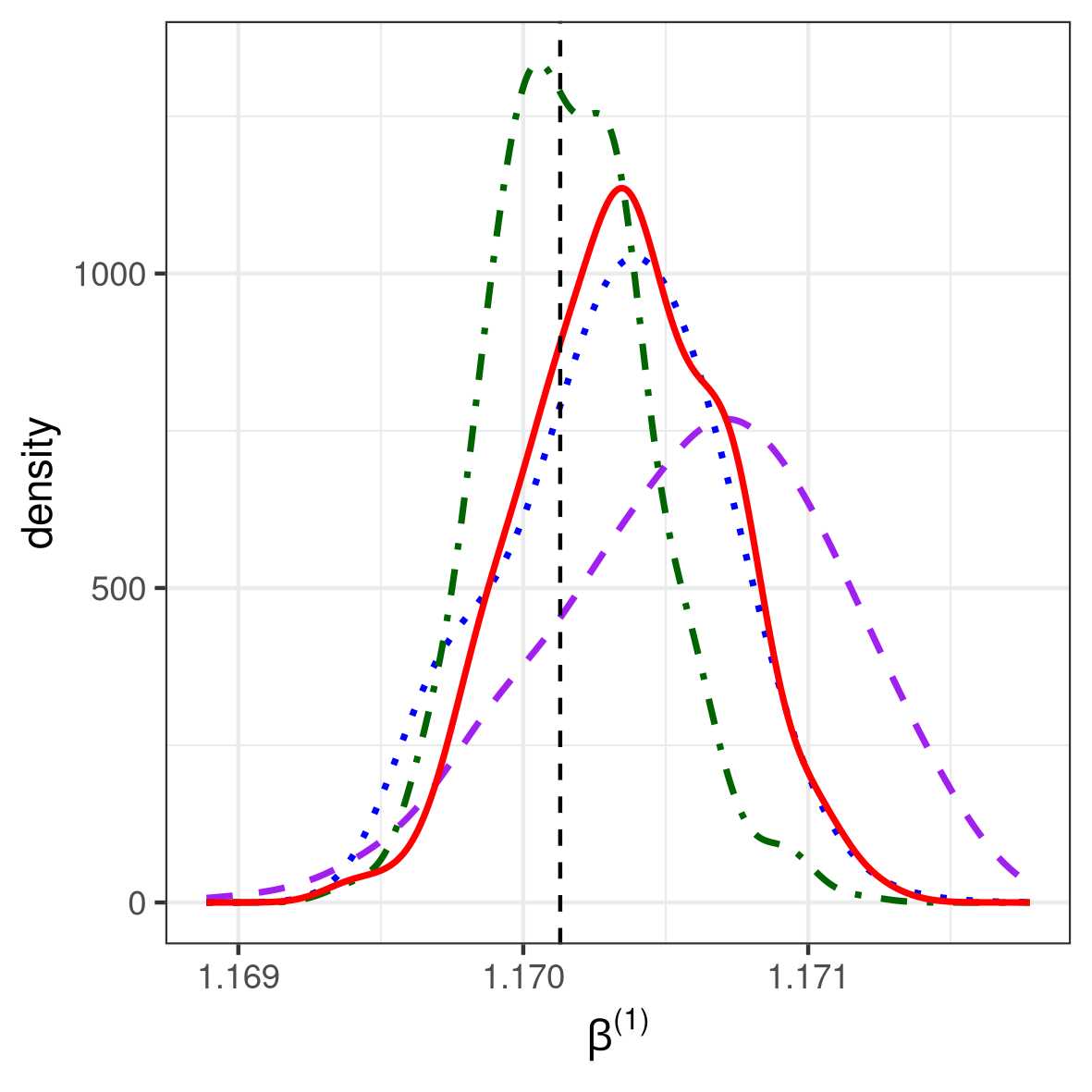}
	\includegraphics[width=0.4\textwidth]{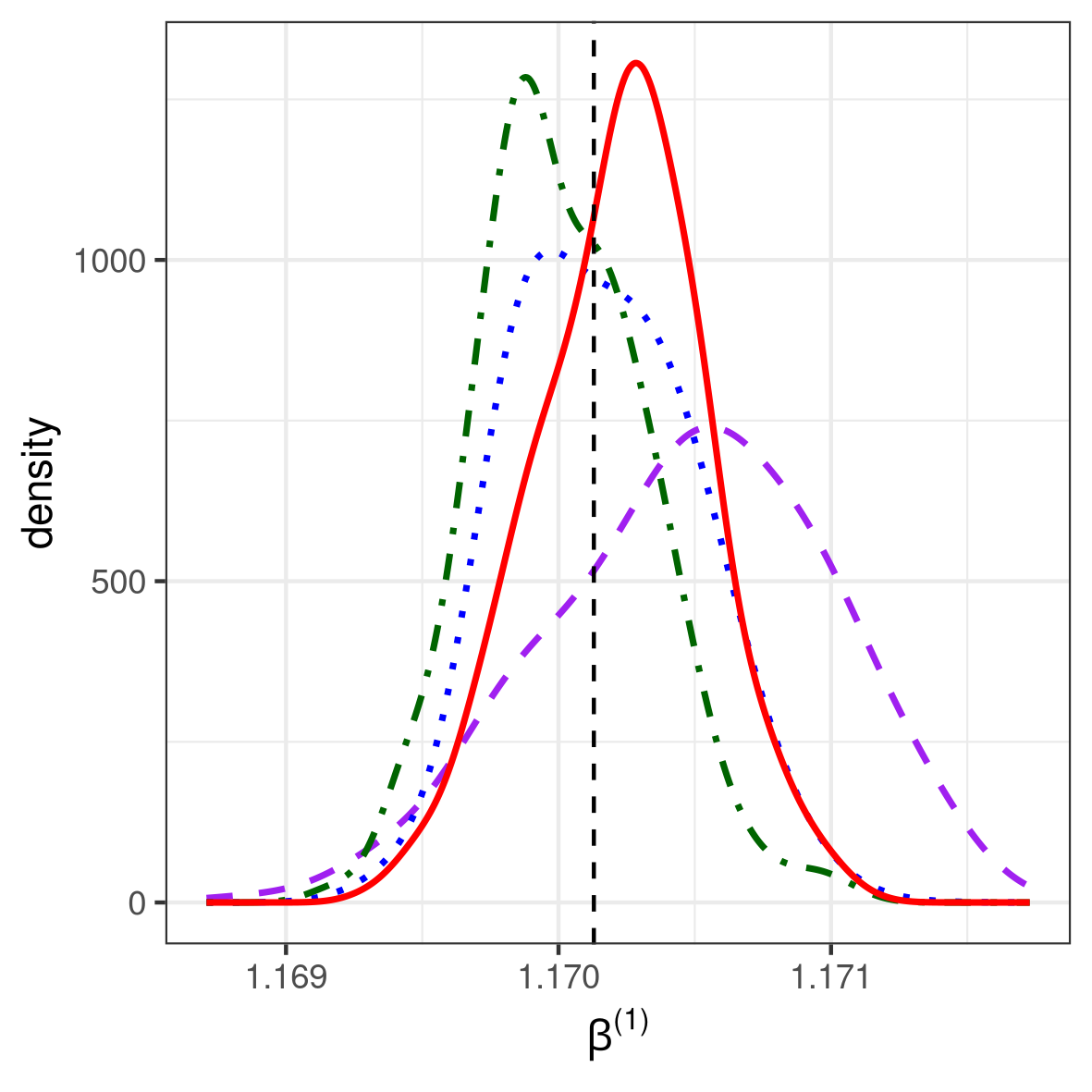}

	\caption{
	Same as right panel of Figure \ref{fig:potts}, but for the other four simulated images.
	}
	\label{fig:potts_2}
\end{figure}

\section{Surrogate Modeling with the Kent Distribution}\label{sec:Kent}

In this section, we give an additional illustration of our proposed methodology with the Kent distribution \citep{kent1982fisher}. This distribution is used to model data on a sphere, and is analogous to a bivariate normal distribution on a Euclidean plane. The likelihood is given by,
\begin{equation}\label{eq:Kent_lik}
p(\zvec \mid \gammavec_1, \gammavec_2, \gammavec_3, \kappa, \beta) = \frac{1}{\mathcal{C}(\kappa, \beta)} \exp{\kappa (\gammavec_1' \zvec) + \beta [(\gammavec_2' \zvec)^2 - (\gammavec_3' \zvec)^2]},
\end{equation}
where $\zvec \in \mathbb{S}^2$, $\gammavec_1$ determines the mean direction, $\gammavec_2$ and $\gammavec_3$ are the major and minor axes, $\kappa$ determines the concentration, and $\beta$ determines the ellipticity ($0 < \beta < \kappa/2$). To simplify the illustration, we have assumed that the parameters $\gammavec_1, \gammavec_2$ and $\gammavec_3$ are fixed and known (in this case to $(1,0,0)', (0,1,0)',$ and $(0,0,1)'$, respectively). 
 We can see from \eqref{eq:Kent_lik} that the Kent distribution is an exponential-family model. In this case the normalizing constant $\mathcal{C}(\kappa, \beta)$ is known; hence, with this model we can easily compare the posterior distribution obtained using our warped gradient-enhanced Gaussian process surrogate model with the true posterior distribution evaluated on a fine discretization of the parameter space.

\begin{figure}[t!]
	\centering
	
	\includegraphics[width=0.8\textwidth]{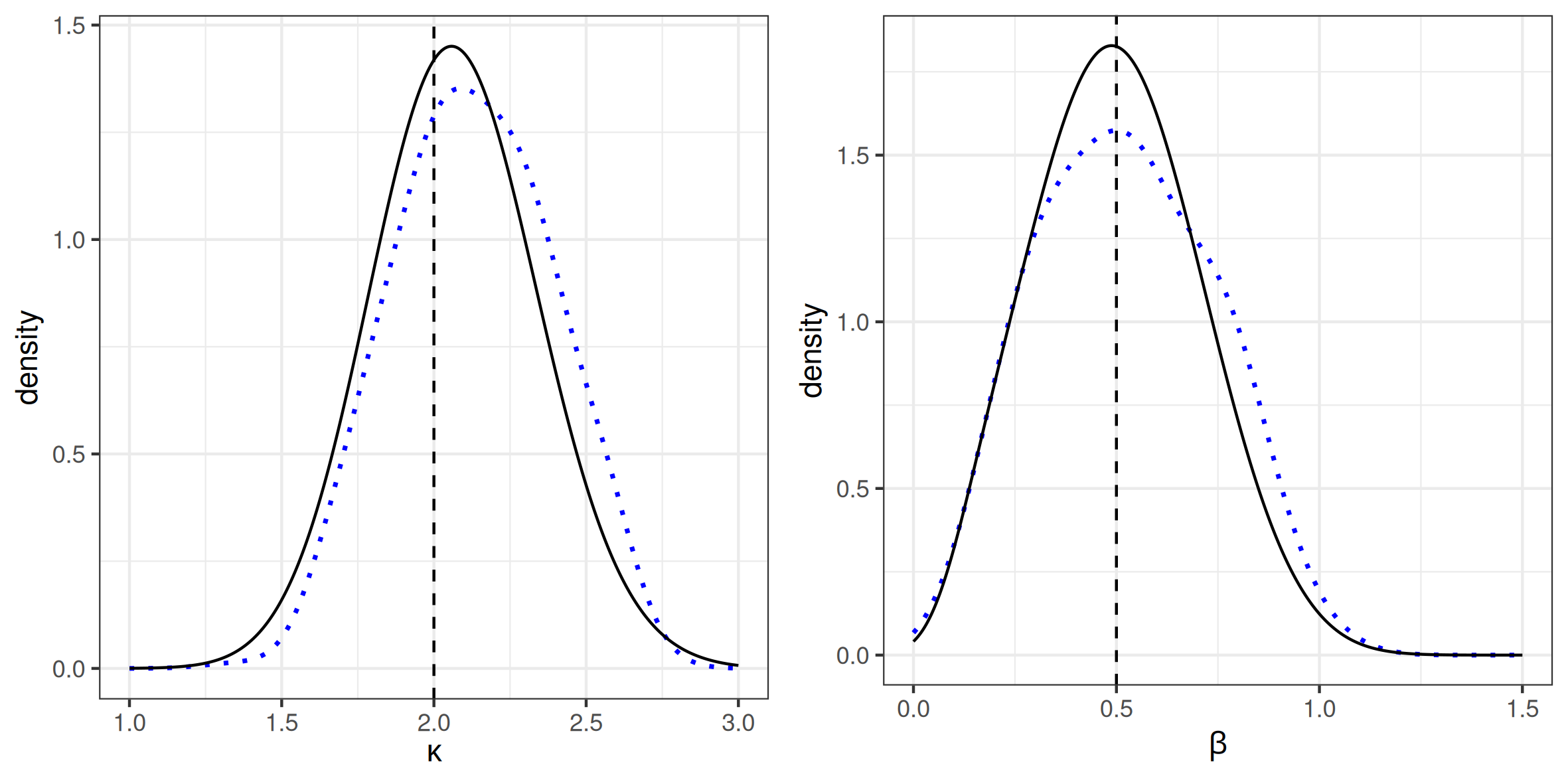}

	\caption{
	Posterior distribution of the parameter $\kappa$ (left panel) and $\beta$ (right panel) for the Kent distribution. Blue dotted line: Importance sampling using the  warped gradient-enhanced Gaussian process surrogate model as proposal distribution. Black solid line: Grid-based approximation to the true posterior distribution. Vertical dashed line: True parameter value.
	}
	\label{fig:kent}
\end{figure}

We performed a simulation study using a sample of 100 data points generated from the Kent distribution with parameters are $\kappa=2$ and $\beta = 0.5$ (chosen close to the ``phase transition'' of the distribution). We used the GE-NS-GP surrogate model to do Bayesian inference using importance sampling with the prior distribution a uniform distribution on the triangular part of the domain $[0.2,10] \times [0.1, 5]$ that lies below the line $\beta = \kappa/2$. To train the GE-NS-GP, we simulated data from the Kent distribution with parameters $\kappa$ and $\beta$ on an equally-spaced grid on this triangular domain. The setup of the surrogate model was otherwise identical to that used in the autologistic model example.
For importance sampling we first took 2000 samples from the surrogate posterior distribution (this was done via a grid approximation), and then simulated sufficient statistics at these 2000 values to obtain a weighted sample. The resulting posterior distribution is shown in Figure \ref{fig:kent}, where we also show the (true) posterior distribution obtained using grid-based methods. As expected, the posterior distributions are very similar. 

\end{document}